\documentclass[preprint,prl,preprintnumbers,amsmath,amssymb]{revtex4}

\usepackage[pdftex]{graphicx}
\usepackage{epstopdf}
\usepackage{dcolumn}
\usepackage{bm}
\usepackage{bibtopic}
\usepackage{color}

\begin{document}


\def\Ef{$E_{\rm F}$}
\def\Ed{$E_{\rm D}$}
\def\Eg{$E_{\rm g}$}
\def\Efmath{E_{\rm F}}
\def\Edmath{E_{\rm D}}
\def\Egmath{E_{\rm g}}
\def\Tc{$T_{\rm C}$}
\def\kpara{${k}_\parallel$}
\def\kparamath{{k}_\parallel}
\def\minuskpara{$-{k}_\parallel$}
\def\kparazero{${k}_{\parallel,0}$}
\def\minuskparazero{$-{\it k}_\parallel,0$}
\def\kperp{${k}_\perp$}
\def\invA{\AA$^{-1}$}
\def\Kbar{$\overline{\rm K}$}
\def\Gbar{$\overline{\Gamma}$}
\def\Mbar{$\overline{\rm M}$}
\def\GbarMbar{$\overline{\Gamma}$-$\overline{\rm M}$}
\def\GbarKbar{$\overline{\Gamma}$-$\overline{\rm K}$}
\def\DeltaEf{$\Delta E_{\rm F}$}
\def\DeltaEfmath{\Delta E_{\rm F}}
\def\BiTe{Bi$_2$Te$_3$}
\def\BiSe{Bi$_2$Se$_3$}
\def\BiMnSe{(Bi$_{1-x}$Mn$_x)_2$Se$_3$}
\def\BiMnTe{(Bi$_{1-x}$Mn$_x)_2$Te$_3$}
\def\BiSnTe{(Bi0.67\%Sn)$_2$Te$_3$}
\def\SbTe{Sb$_2$Te$_3$}
\def\MS{\textcolor{red}{}}
\def\Tc{$T_{\rm C}$}

\title{Nonmagnetic band gap at the Dirac point of the magnetic\\ topological insulator \BiMnSe}

\author{J. S\'anchez-Barriga,$^1$ A. Varykhalov,$^1$ G. Springholz,$^2$ H. Steiner,$^2$ R. Kirchschlager,$^2$ G. Bauer,$^2$ O. Caha,$^3$ E. Schierle,$^1$ E. Weschke,$^1$ A. A. \"Unal,$^1$ S. Valencia,$^1$, M. Dunst,$^5$ J. Braun,$^5$ H. Ebert,$^5$ J. Min\'ar,$^{5,6}$, E. Golias,$^1$ L. V. Yashina,$^7$ A. Ney,$^2$ V. Hol\'y,$^4$ O. Rader$^1$}

\affiliation{$^1$Helmholtz-Zentrum Berlin f\"ur Materialien und Energie, 
Elektronenspeicherring BESSY II, Albert-Einstein-Stra\ss e 15, 12489 Berlin, Germany\\
$^2$Institut f\"ur Halbleiter- und Festk\"orperphysik, Johannes Kepler Universit\"at, Altenbergerstr. 69,  4040 Linz, Austria\\ 
$^3$CEITEC and Department of Condensed Matter Physics, Masaryk University, Kotlarska 2, 61137 Brno, Czech Republic\\ 
$^4$Department of Condensed Matter Physics, Charles University, Ke Karlovu 5, 12116 Prague, Czech Republic\\ 
$^5$Department Chemie, Ludwig-Maximilians-Universit\"at M\"unchen, Butenandtstr. 5-13, 81377 M\"unchen, Germany\\ 
$^6$New Technologies Research Centre, University of West Bohemia, Univerzitni 8, 306 14 Pilsen, Czech Republic\\ 
$^7$Department of Chemistry, Moscow State University, Leninskie Gory 1/3, 119991, Moscow, Russia}

\begin{abstract}

Magnetic doping is expected to open a band gap at the Dirac point of topological insulators by breaking time-reversal symmetry and to enable novel topological phases. Epitaxial (Bi$_{1-x}$Mn$_{x}$)$_{2}$Se$_{3}$ is a prototypical magnetic topological insulator with a pronounced surface band gap of $\sim100$ meV. We show that this gap is neither due to ferromagnetic order in the bulk or at the surface nor to the local magnetic moment of the Mn, making the system unsuitable for realizing the novel phases. We further show that Mn doping does not affect the inverted bulk band gap and the system remains topologically nontrivial. We suggest that strong resonant scattering processes cause the gap at the Dirac point and support this by the observation of in-gap states using resonant photoemission. Our findings establish a novel mechanism for gap opening in topological surface states which challenges the currently known conditions for topological protection.


\end{abstract} 

\maketitle

The topological surface state (TSS) of the three-dimensional topological insulator (3D TI) \BiSe\ forms a Dirac cone in the band dispersion $E({k}_\parallel)$  of electron energy $E$ versus wave vector component ${k}_\parallel$ parallel to the surface \cite{Has10}. Differently from graphene, the electron spin is nondegenerate and locked to ${k}_\parallel$, and the Dirac crossing point is protected by time-reversal symmetry (TRS) against distortions of the system \cite{Has10,Qi11}. This stability has been theoretically investigated demonstrating that nonmagnetic impurities at the surface and in the bulk can form a resonance in the surface-state density of states (DOS) without opening a band gap \cite{LeePRB09,BiswasPRB10,LuNJP11}. A magnetic field breaks TRS lifting the Kramers degeneracy $E({k}_\parallel,\uparrow) = E(-{k}_\parallel,\downarrow)$ between up ($\uparrow$) and down ($\downarrow$) spins, and can open a band gap at the Dirac point \cite{Has10}. For the two-dimensional quantum-spin-Hall system HgTe, the effect of a perpendicular magnetic field on the topologically-protected edge state has been demonstrated successfully \cite{KoenigScience2007}, and similar effects are expected from magnetic impurities in this system \cite{CXLiuPRL08}. In subsequent studies on HgTe, the effect of the magnetic field was much smaller, and recently in an inverted electron-hole bilayer from InAs/GaSb, helical edge states proved robust in perpendicular magnetic fields of 8 T \cite{Dupreprint}. At the surface of a 3D TI, calculations show that  magnetic impurities can open a gap at the Dirac point and exhibit ferromagnetic order with perpendicular anisotropy mediated by the TSS through Ruderman-Kittel-Kasuya-Yosida (RKKY) exchange coupling \cite{Liu09,BiswasPRB10,WrayNatPhys2011}.

3D TIs with magnetic impurities have been studied by angle-resolved photoelectron spectroscopy (ARPES). The magnetic impurities have been employed in the bulk \cite{ChenScience2010} and at the surface of \BiSe\ \cite{WrayNatPhys2011}. For Fe impurities, the opening of large surface band gaps of $\sim$100 meV was reported in both cases \cite{ChenScience2010,WrayNatPhys2011}. In a previous work, however, we found that Fe deposited on \BiSe\ does not open a gap for a wide range of preparation conditions, revealing a surprising robustness of the TSS towards magnetic moments \cite{Scholz}. This conclusion also extends to Gd/\BiSe\ \cite{Valla} and Fe/\BiTe\ \cite{ScholzPSS13}. 

The absence of a surface band gap is consistent with the recent finding that Fe on \BiSe\ does not favor a magnetic anisotropy perpendicular to the surface, at least in the dilute limit \cite{HonolkaPRL12}. In this respect, bulk impurities in TIs have an advantage over surface impurities in that ferromagnetic order and perpendicular magnetic anisotropies have been achieved in the bulk systems \cite{ChenScience2010,Kulbachinskii01,Kulbachinskii02,Dyck02,HorPRB2010,XuNatPhys12,Checkelsky12}. Fe incorporated in bulk \BiTe\ is known to order ferromagnetically with a Curie temperature ($T_{\rm C}$) of $\sim$12 K for concentrations of $x=0.04$ showing an easy axis perpendicular to the base plane \cite{Kulbachinskii01,Kulbachinskii02}. On the other hand, (Bi$_{1-x}$Fe$_x)_2$Se$_3$ has not been found to be ferromagnetic at temperatures above 2 K \cite{Kulbachinskii02}.
For $x=0.16$ and $0.25$, ferromagnetic order was found at 2 K \cite{ChenScience2010}. By ARPES \cite{Hufner}, surface-state band gaps in (Bi$_{1-x}$Fe$_x$)$_2$Se$_3$ were reported and assigned to a magnetic origin for Fe concentrations from $x=0.05$ to $0.25$ including nonferromagnetic concentrations such as $x=0.12$ with a band gap of 45 meV \cite{ChenScience2010}. These band gaps have been attributed to short-range magnetic order. Concerning Mn incorporation, bulk \BiMnSe\ with $x=0.02$ has been shown to exhibit a surface band gap with an occupied width of 7 meV \cite{ChenScience2010}, which was suggested as indication of ferromagnetic order induced by the TSS \cite{ChenScience2010}. However, much larger surface band gaps were observed for $n$-doped \BiMnSe\ films \cite{XuNatPhys12}, where ferromagnetic order of Mn at the surface was found to be strongly enhanced with $T_{\rm C}$ up to $\sim$45 K. The ferromagnetic order resulted in an unusual spin texture of the TSS. The $T_{\rm C}$ at the surface was much higher than in the bulk \cite{XuNatPhys12}, which was partially attributed to Mn surface accumulation. A strong enhancement of the surface $T_{\rm C}$ was also predicted by mean-field theory for this system [e.g., from 73 K (bulk) to 103 K (surface)] \cite{RosenbergPRB12}. Magnetically-doped TIs with ferromagnetic order are important as realizations of novel topological phases. When spin degeneracy is lifted by the exchange splitting, bulk band inversion can occur selectively for one spin. If also the Fermi level is in the band gap, as predicted for Cr and Fe in \BiSe\ \cite{QAHE2}, this gives rise to an integer quantized Hall conductance $\sigma_{xy}$ in thin films termed the quantized anomalous Hall effect \cite{CXLiuPRL08,QAHE2}. This has recently been reported for Cr in (Bi,Sb)$_{2}$Te$_{3}$ \cite{QAHE3}. When a perpendicular magnetization breaks TRS at the surface of a bulk TI, the resulting mass and surface band gap give rise to quantized edge states. In this case $\sigma_{xy}$ is half-integer quantized yielding a half quantum Hall effect \cite{QiPRB08} which can be probed at ferromagnetic domain boundaries at the surface \cite{Pankratov87} and leads to exotic topological magnetoelectric effects such as the point-charge-induced image magnetic monopole \cite{QiScience09,ZangNagaosaPRB10}. Another topological phase predicted for magnetically doped \BiSe\ is the realization of a 3D Weyl fermion system in which the Dirac-like dispersions become a property of the bulk \cite{Weyl}.  

Here we present a detailed investigation on the correlation between the TSS and magnetic properties in the Bi-Mn-Se system, where the magnetic impurity Mn is introduced in the bulk of $\sim$0.4 $\mu$m thick \BiMnSe\ epilayers with concentrations up to $x=0.08$. By ARPES we find remarkably large band gaps of $\sim$200 meV persisting up to temperatures of 300 K far above $T_{\rm C}$, which is found to be $\sim$10 K at the surface showing only a limited enhancement by $\le 4$ K over the bulk $T_{\rm C}$. We find that the magnitude of the band gaps by far exceeds that expected from TRS breaking due to magnetic order. By ARPES from quantum-well states we exclude that Mn changes the inverted bulk band gap. Using resonant photoemission and {\it ab-initio} calculations we conclude that, instead, impurity-induced resonance states destroy the Dirac point of the TSS. We further support our conclusion by showing that extremely low bulk concentrations of nonmagnetic In do also open a surface band gap in \BiSe. Our findings have profound implications for the understanding of the conditions for topological protection.\\

\noindent{\bf RESULTS}

\noindent {\bf Concentration and temperature dependence of surface band gap.}
Figure 1 shows Mn-concentration-dependent high-resolution ARPES dispersions of the TSS and bulk valence band (BVB) states of \BiMnSe\ with $x$ up to $0.08$ measured at 12 K and 50 eV photon energy. At this energy, contributions from the bulk-conduction band (BCB) do not appear due to the dependence of the photoemission transitions on the component of the electron wave vector perpendicular to the surface ${k}_\perp$ \cite{Hufner}, but the BCB is partially occupied as data at 21 eV photon energy reveal (Supplementary Fig. 1). The clean \BiSe\ film (Fig. 1a) is $n$-doped and exhibits a well-resolved Dirac point with high photoemission intensity at a binding energy of \Ed$\sim$0.4 eV that is seen as an intense peak in the energy-distribution curve (EDC) at zero momentum superimposed as red curve on the right-hand side of the panel. The intact and bright Dirac point marked by a horizontal dashed line in Fig. 1a demonstrates that the TSS is gapless in \BiSe. At larger binding energies, the lower half of the Dirac cone overlaps with the BVB, which is observed for all \BiMnSe\ samples (Figs. 1b-d). Increasing Mn concentration, we find a gradual upward shift of the band edges in energy, revealing a progressive $p$-type doping (hole doping). Most notably, a surface band gap opens at the Dirac point with increasing Mn content. In each EDC, the opening of the gap is also evident from the development of an intensity dip at the binding energy of the original Dirac point. Thus, the energy separation between the upper Dirac band minimum and the energy position of the intensity dip is about half of the gap size. Note that the horizontal dashed lines in Figs. 1b-d highlight the minimum limit of the gap size after taking into account the contribution from the linewidth broadening to the ARPES spectra (see Supplementary Note 1 and Supplementary Fig. 2). The surface band gap rapidly increases with Mn content and exceeds 200 meV for $x= 0.08$. For $x=0.02$, the suppression of the Dirac point intensity as compared to the undoped \BiSe\ film indicates the existence of a small gap. This is consistent with the fact that the TSS dispersion becomes slightly parabolic and is characterized by an increased effective mass of $m^*\approx 0.09 m_e$ for $x=0.02$. This value increases further to $\approx 0.15 m_e$ for $x=0.08$, where $m_e$ is the free-electron mass.

The observed surface band gaps are similar to those previously reported \cite{XuNatPhys12}, which have been attributed to long-range ferromagnetism and TRS breaking of the topologically-protected surface
state \cite{XuNatPhys12}. Figures 2a-d show high-resolution ARPES dispersions of the TSS as well as the normal-emission spectra through the Dirac point at temperatures of 12 and 300 K for $x=0.08$ (see also Supplementary Fig. 3). Strikingly, there is no significant change as temperature is raised, and very clearly the band gap in the TSS persists up to room temperature. Moreover, we find a similar behavior for the whole sample series independently of the Mn content, which challenges the dominant role of ferromagnetic order in inducing the band gap in the TSS, unless the surface $T_{\rm C}$ is above room temperature.
\\

\noindent {\bf Bulk and surface magnetism.}
We begin in Fig. 3 with characterizing the bulk magnetic properties using 
superconducting quantum-interference device (SQUID) magnetometry. The comparison of measurements with in-plane and out-of-plane applied magnetic field in Fig. 3a shows that at a temperature of 2 K the bulk easy axis lies in the surface plane. This holds irrespectively of the Mn concentration. Since the opening of a gap at the Dirac point requires a magnetization perpendicular to the surface \cite{Liu09}, we concentrate in the following on the out-of-plane component of the magnetization. In the hysteresis loops displayed in Fig. 3b measured with an out-of-plane applied magnetic field at different temperatures, we observe a narrow magnetic hysteresis at 4.2 K, whereas paramagnetic behavior emerges at 7 K once the ferromagnetic transition has been crossed (inset in Fig. 3b). For a better determination of $T_{\rm C}$ we present in Fig. 3c modified Arrott plots according to a 3D Heisenberg ferromagnet, normalized to the mass of the sample, with exponents $\beta = 0.348$ and $\gamma= 1.41$, from which we deduce $T_{\rm C}=5.5$ K in the bulk (see also Supplementary Fig. 4). This is well below the temperature of our ARPES measurements shown in Figs. 1 and 2. 

The only remaining possibility that long-range ferromagnetism opens the gap at the Dirac point is an enhanced surface magnetism with magnetization perpendicular to the surface \cite{Liu09, XuNatPhys12, RosenbergPRB12}. This cannot be verified directly by the bulk-sensitive SQUID magnetometry. Therefore, we perform x-ray magnetic circular dichroism (XMCD) measurements at the Mn $L_{2,3}$-edges to probe the near-surface ferromagnetic order. The detection by total electron yield leads to a probing depth in the nanometer range. Figure 4a shows for $x=0.04$ the normalized intensity of the Mn $L_{2,3}$ absorption edges obtained upon reversal of the photon helicity at a temperature of 5 K with an out-of-plane applied magnetic field of 3 T. The XMCD difference spectrum is shown in Fig. 4b, with the normalized XMCD difference signal following one part of the out-of-plane hysteresis as a function of the applied magnetic field as inset. The temperature dependence of the XMCD signal measured in remanence (0 T applied magnetic field) is presented in Fig. 4c for Mn concentrations of $x=0.04$ and 0.08. This signal is obtained after switching off an applied magnetic field of 5 T perpendicular to the surface, and represents the remanent XMCD, which is proportional to the remanent magnetization of the film surface. We clearly observe a ferromagnetic Mn component in the remanent XMCD appearing around 640 eV which becomes increasingly more pronounced at lower temperatures. The lineshape of the XMCD spectrum is similar to that of Mn in GaAs \cite{GaMnAs}, indicating a predominant $d^5$ configuration. More specifically, the lineshape compares rather well with Mn in \BiTe\ for which the comparison to an Anderson impurity model recently gave a superposition of 16\%\ $d^4$, 58\%\ $d^5$ and 26\%\ $d^6$ character \cite{Watson}. The ferromagnetic Mn component shows surface ferromagnetic order in the remanent XMCD only below 10 K for Mn concentrations of $x=0.04$ and 0.08 (inset in Fig. 4c). On the one hand, this result is not inconsistent with the predicted strong surface enhancement of $T_{\rm C}$ since the enhancement effect vanishes already when chemical potential and Dirac point differ by 0.2 eV \cite{RosenbergPRB12}. On the other hand, our XMCD result means that the magnetically induced gap would have to close above 10 K.
 
Having established that the bulk and surface magnetic properties of our samples are very similar, as a next step, we performed field-cooling experiments to re-examine the bulk magnetic properties. Figure 4d shows field-cooled SQUID data for an applied magnetic field of 10 mT parallel and perpendicular to the surface plane. Above $T_{\rm C}$, there is no preferential orientation of the anisotropy axis perpendicular to the surface. Additional zero-field-cooled SQUID data compare rather well with these results (Supplementary Fig. 5). Further XMCD measurements under an in-plane and out-of-plane applied magnetic field are consistent with the SQUID data and contained in Supplementary Fig. 6. These XMCD and SQUID results strongly suggest that above $T_{\rm C}$ also short-range static magnetic order does not play a role at the surface or in the bulk, respectively (see Supplementary Note 2), and that domains with partial or full out-of-plane magnetic order are absent. We further support this conclusion by additional XMCD measurements carried out by means of x-ray photoelectron emission microscopy (XMCD-PEEM) at room temperature, with a lateral resolution of $\sim$20 nm. Figure 4e shows the XMCD image which due to the light incidence (16$^\circ$ grazing) is sensitve to both in- and out-of-plane components of the magnetization, with additional data given in Supplementary Figs. 7,8. The XMCD image displays no magnetic domains at room temperature thus also excluding short-range static inhomogeneous magnetic order with magnetization partially or fully in or out of the surface plane, at least within our lateral resolution (see Supplementary Note 3 for details). A similar conclusion can be drawn regarding out-of-plane magnetic short-range fluctuations (see discussion in Supplementary Note 4 and Supplementary Fig. 9). All our observations taken together lead us to the important conclusion that the surface band gap is not due to magnetism and, thus, the novel topological phases cannot be realized with \BiMnSe. 
 \\

\noindent {\bf Absence of topological phase transition with concentration.}
There exists an alternative nonmagnetic explanation for the surface band gap. 
It is in principle possible that the incorporation of Mn changes the bulk band structure and even reverts the bulk band inversion, rendering Mn-doped \BiSe\ topologically trivial. This topological phase transition with concentration is, e.g., the basis for the HgTe quantum spin Hall insulator  \cite{KoenigScience2007}. Interestingly, it has been argued that a gap in the TSS can be a precursor of a reversed bulk band inversion, as discussed for 
TlBi(S$_{1-x}$Se$_x$)$_2$ \cite{SatoNP11}. Indium substitution in \BiSe\ behaves similarly, and leads to a topological-to-trivial quantum phase transition with an inversion point between 3\% and 7\% In in thin films \cite{HasanBiInSe}. Figure 5 shows that the bulk band gap stays constant within 4\%\ of its value for 8\%\ Mn incorporation. This statement is possible because at low photon energies changes in the bulk band gap are traced most precisely from quantum-well states in normal emission (${k}_\parallel$=0). The simultaneous quantization of BCB and BVB is created by surface band bending after adsorption of small amounts of residual gas \cite{BianchiPRL11}. This means that the scenario of reversed bulk band inversion does not hold here. 

Another remarkable feature is the measured size of the surface band gap. 
A perpendicular magnetic anisotropy has recently been predicted by density-functional theory (DFT) for Co in \BiSe\ \cite{Schmidt} as well as for 0.25 monolayer Co adsorbed on the surface in substitutional sites, leading to a surface band gap of $\sim9$ meV \cite{Schmidt} which, however, is one order of magnitude smaller than the gaps observed here. For \BiMnTe, a surface band gap of 16 meV has been calculated under the condition of perpendicular magnetic anisotropy \cite{HenkPRL12}. For Mn in \BiSe\ at the energetically favored Bi-substitutional sites, an in-plane magnetic anisotropy is predicted \cite{Abdalla}. The size of the gap at the Dirac point is only 4 meV \cite{Abdalla}. Perpendicular anisotropy and ferromagnetic order are absent in our samples at the temperature of the ARPES measurements.
\\

\noindent {\bf Role of the local magnetic moment and intercalation scenarios.}
If the large local magnetic moment of the Mn is responsible for the measured band gaps via TRS breaking, the effect should be equal or larger when Mn is deposited directly on the surface. Here we argue in the following way: If TRS is broken only due to the Mn magnetic moment, we do not require hybridization to open a gap. By depositing Mn we create a completely different environment for the Mn at the surface, allowing us to understand whether the Mn magnetic moment is solely responsible for the opening of large surface band gaps via TRS breaking and, ideally, without hybridization playing a role. Mn is very well suited for such comparison, as due to the high stability of its half-filled $d^5$ configuration, its high magnetic moment is to a large extent independent of the atomic environment and the resulting hybridization. DFT obtains 4.0 $\mu_{\rm B}$ for most substitutional sites (except for the hypothetical Se-substitutional site with still 3.0 $\mu_{\rm B}$) and only for interstitial Mn the moment peaks with 5.0 $\mu_{\rm B}$ \cite{Abdalla}. Mn deposition has been performed at $\sim$30 K in order to keep the Mn atoms isolated from each other and on the surface. As the XMCD showed a predominant $d^5$ configuration for Mn bulk impurities, which is most stable, we can assume the same magnetic moment for Mn impurities deposited on the surface. Figure 6 shows that a similar $p$-doping occurs as with the Mn in the bulk. In another respect, the magnetic moments do not act in a similar way as in the bulk. Even 30\%\ of a Mn monolayer on \BiSe\ does not open a band gap at the Dirac point. This result is similar to what we have observed for Fe on \BiSe\ \cite{Scholz} and \BiTe\ \cite{ScholzPSS13}.

Most recently, the appearance of surface band gaps in ARPES at the Dirac point of the TSS has been discussed based on different mechanisms. One is momentum fluctuations of the surface Dirac fermions in real space as observed by scanning tunneling microscopy measurements of Mn-doped \BiTe\ and \BiSe\ bulk single crystals \cite{Beidenkopf-NatPhys-2011}. However, these fluctuations amount to 16 meV \cite{Beidenkopf-NatPhys-2011}, much less than the present band gaps. Another mechanism is top-layer relaxation, i.e., breaking of the van-der-Waals bond between quintuple layers during cleavage of bulk crystals as proposed in Ref. \cite{XuNatPhys12}. This would open a gap due to hybridization of TSS's through the layer \cite{YZhangNP10} but such an effect does not occur for epitaxial layers where no sample cleavage is employed for surface preparation. 

If such a separation of quintuple layers occurs, it is more likely caused by the Mn. Principally, intercalated Mn in the van-der-Waals gaps could separate
quintuple layers electronically although such effect has not been seen in experiments yet. At these new interfaces TSS's could form and if they would behave like in ultrathin \BiSe\ films, they would hybridize and open a band gap \cite{YZhangNP10}. However, band gaps of the order of $\sim$100 meV would correspond to an unrealistic Mn intercalation pattern grouping three quintuple layers when compared to the films of Ref. \cite{YZhangNP10}. In order to estimate the amount of Mn intercalation, we have analyzed the change of the bulk lattice constant in our samples by x-ray diffraction (see Supplementary Note 5, Supplementary Figs. 10-13 and Supplementary Tables 1,2). We find that the $c$-lattice constant increases by $\sim 0.15$\%\ for 4\%\ Mn and by $\sim 0.45$\%\ for 8\%\ Mn. If we would assume that this is completely due to an expansion of the van-der-Waals gaps, the interlayer distance would increase by only $\sim 1.8$\%\ for 4\%\ Mn and by less than $\sim 5$\%\ for 8\%\ Mn. According to DFT, intercalation of transition metals into the first van-der-Waals gap of \BiSe\ leads to an expansion between 10 and 20\%\ \cite{Eremeev12}. New surface-state features can appear inside the bulk band gap but in order to be split remarkably off the BVB and BCB edges, expansions between 20 and 50\%\ are required \cite{Eremeev12}. In addition, we point out that, despite our observation of a large surface band gap at room temperature, we observe a nearly-zero spin polarization perpendicular to the surface (Supplementary Fig. 14). The absence of a hedgehog spin texture pinpoints a non-TRS breaking mechanism underlying the origin of the gap, a fact which is consistent with our conclusions derived from the magnetic properties. Moreover, it is understood that we cannot verify a small magnetic band gap opening of few meV as calculated in Refs. \onlinecite{Schmidt,HenkPRL12} because much larger gaps are present in the whole temperature range.\\

\noindent {\bf Topological protection beyond the continuum model.}
As we find that neither ferromagnetic order in the bulk or at the surface, nor the local magnetic moment of the Mn are causing the large band gaps that we observe, as we can exclude the nonmagnetic explanation of a reversal of the bulk band inversion and as we do not find sufficient indications for the nonmagnetic explanation of surface-state hybridization by van-der-Waals-gap expansion, 
we point out a different mechanism based on impurity-induced resonance states that locally destroy the Dirac point. 

In fact, recently the treatment of the topological protection in the continuum model \cite{LeePRB09,BiswasPRB10,LuNJP11} has been extended for finite bulk band gaps \cite{BlackSurface12,BlackBulk12}. As a result, surface \cite{BlackSurface12} and bulk impurities \cite{BlackBulk12} can mediate scattering processes via bulk states and the localized impurity-induced resonance states emerging at and around the Dirac point lead to a local destruction of the topological protection of the Dirac point. The resulting band gap opening depends on the resonance energy, impurity strength $U$ as well as spatial location of the impurities
\cite{BlackSurface12,BlackBulk12}. A typical gap size is of the order of 100 meV \cite{BlackSurface12}. Moreover, the gap does not rely on a particular magnetic property of the impurity and, therefore, the mechanism should apply for magnetic impurities as well. The model can also explain the absence of the gap when Mn impurities are placed on the surface, since bulk-like resonance states do not form when Mn is placed only on the surface. A definite assignment, however, requires realistic electronic structure calculations since a simple atomic Mn $d^5$ configuration does not lead to states at the Dirac point. 

To identify impurity resonances, we calculated for 8\%\ Mn in \BiSe\ at the Bi-substitutional sites the corresponding DOS by {\it ab-initio} theory. The model structure used in the calculations is shown in Fig. 7a, where Bi atoms (yellow) acquire Mn character (blue). The results of the calculation are shown in Fig. 7b, where it is seen that Mn impurity states (blue) strongly contribute to the total DOS (red) near the BVB maximum. The Mn impurity states in the gap are clearly identified by assuming ferromagnetic order and subtracting the minority from majority spin DOS. In experiment, impurity-band states are difficult to observe as the example of Ga$_{1-x}$Mn$_x$As teaches \cite{Gray12}, where similarly to the present case impurity states emerge near the BVB maximum. There are, in fact, indications for the impurity resonances in our data. If we look at the photon-energy dependence from 16 to 21 eV in the ARPES results shown in Supplementary Fig. 15, we see that the surface band gap is only well defined with respect to the minimum of the upper Dirac cone and consistent for 16--21 eV as well as 50 eV (for 50 eV see Figs. 1,2 where the BCB does not appear). The lower Dirac cone exhibits despite its two-dimensional nature a photon-energy (wave vector perpendicular to the surface, ${\it k}_\perp$) dependence when Mn is incorporated. The reason for this difference is that the lower Dirac cone strongly overlaps with the BVB while the upper Dirac cone does not overlap with the BCB. This can also be seen in Fig. 5, where an apparent band gap of $\sim200$ meV occurs for 2\%\ Mn at $h\nu=16$ eV. At this small concentration, such an apparent gap seems to be nearly closed at 18 eV, where the suppression of the Dirac point intensity and the more parabolic TSS dispersion as compared to the undoped \BiSe\ film indicates the existence of a small gap, in agreement with the results of Fig. 1. Note that the minimum band gap in the photon-energy dependence defines the gap size (see Supplementary Fig. 16). The minimum gap size observed at low photon energies agrees well with the one obtained at 50 eV. At 8\%\ Mn, the surface band gap opens at all photon energies and reaches a minimum of $\sim200$ meV, i.e., it is determined rather unambiguously, but some intensity appears in the surface band gap. Such intensity also appears in Ref. \onlinecite{XuNatPhys12}. The role of the impurity resonance is exactly to couple the TSS to bulk states. This 3D coupling is naturally different for the upper and lower Dirac cone due to the different bulk overlap and thus seen as the photon-energy dependence of the lower half of the Dirac cone. 

We applied resonant photoemission which fortunately is comparatively strong for Mn due to its half-filled $d$-shell. Figures 7c,d show for 8\%\ Mn off-resonant ($h\nu=48$ eV) and on-resonant photoemission (50 eV) measurements via the Mn 3$p$ core level, focusing on the region of the surface gap, respectively. The spectra were normalized to the photon flux after taking into account the photon energy dependence of the photoionization cross sections of Bi 6$p$ and Se 4$p$, respectively. Subtracting off-resonant from on-resonant data allows us to visualize directly the contribution from impurity resonances in the ARPES spectrum, as there is enhanced photoemission from d-like Mn states through decay of electrons that are excited via 3$p$-3$d$ transitions (Fig. 7e). The difference spectrum shown in Fig. 7f reveals the existence of impurity resonances inside the surface band gap and near the valence-band maximum. The resonances are seen in blue contrast and are marked with arrows, similar to the calculations shown in Fig. 7b. The unexpectedly strong dispersion of the resonant states with wave vector ${k}_\parallel$ is in qualitative agreement with our one-step model photoemission calculations (see Supplementary Fig. 17 and Supplementary Note 6), and additionally supports the physical picture of Mn-induced coupling to the bulk states. We should emphasize that the Mn $d^5$ configuration confirmed by our XMCD measurements offers much fewer states in the energy range of the Dirac cone than the other magnetic transition metals. The fact that already Mn breaks the Dirac cone indicates that the present result is of general importance for TIs doped with magnetic transition metals. 

As nonmagnetic control experiment we have chosen thick In-doped \BiSe\ bulk samples which give rise to a trivial phase close to $\sim5$\%\ In concentration. Supplementary Fig. 18 shows that a large band gap of the order of $\sim100$ meV appears at the Dirac point already for a much smaller In concentration, namely 2\%, far away from the inversion point of the topological quantum-phase transition (see Supplementary Note 7 and Supplementary Fig. 19 for more details). In addition, spin-resolved ARPES measurements around ${k}_\parallel=0$ for In-doped samples (Supplementary Fig. 14) reveal that the out-of-plane spin polarization is zero, similarly to the result for the gapped Dirac cone in Mn-doped samples. Interestingly, the size of the band gap for 2\%\ In is of the same order as the one for 8\%\ Mn whereas no gap appears for 4\%\ Sn (Supplementary Fig. 18). This indicates that the concentration range at which the large surface gap develops varies from dopant to dopant. Based on the ideas proposed recently \cite{BlackSurface12,BlackBulk12}, this might be associated with the impurity-dependent strength $U$, regardless whether the dopant is magnetic or not. Our control experiments demonstrate the existence of a novel mechanism for surface band gap opening which is not directly connected to long-range or local magnetic properties. Although it is not possible to directly search for In impurity resonances in the photoemission experiment as there is no resonant photoemission condition available, we point out for completeness that our conclusion on the nonmagnetic origin of the surface band gap in Mn-doped \BiSe\ films would not have been possible unless several findings in our experiment differed from previous experiments \cite{XuNatPhys12}. This refers to the in-plane magnetic anisotropy, the absence of a temperature dependence of the gap, a much lower $T_{\rm C}$, no out-of-plane spin polarization at the gapped Dirac point, no enhanced surface magnetism, a photon-energy dependence of the band gap, and an available resonant condition via the Mn 3$p$ core level allowing us to directly observe the contribution from in-gap states.   

To summarize, we revealed the opening of large surface band gaps in the TSS of Mn-doped \BiSe\ epilayers that strongly increase with increasing Mn content. We find ferromagnetic hysteresis at low temperatures, with the magnetization oriented parallel to the film surface. Magnetic surface enhancement is absent. Even above the ferromagnetic transition, we observe surface band gaps which are one order of magnitude larger than the magnetic gaps theoretically predicted and, moreover, they do not show notable temperature dependence. No indication for a Mn-induced reversal of the bulk band inversion and no significant enhancement of the surface magnetic ordering transition with respect to the bulk are found. Control experiments with nonmagnetic bulk impurities, conducted in the topological phase, reveal that surface band gaps of the order of $\sim$100 meV can be created without magnetic moments. In line with recent theoretical predictions, we conclude that the band gap opening up to room temperature in Mn-doped films is not induced by ferromagnetic order and that even the presence of magnetic moments is not required. Our results are important in the context of topological protection and provide strong circumstantial evidence that Mn-doped \BiSe\ is not suited for observing the quantized anomalous Hall effect or the half quantum Hall effect.\\ 
 
\noindent{\bf METHODS}

\noindent {\bf Sample growth and structural characterization.}
The samples were grown by molecular beam epitaxy on (111)-oriented BaF$_2$ substrates at a substrate temperature of $360\,^{\circ}{\rm C}$ using \BiSe, Mn, and Se effusion cells. The Mn concentration was varied between 0 and 8\%, and a two-dimensional growth was observed by {\it in situ} reflection high-energy electron diffraction in all cases (see Supplementary Fig. 20 and Supplementary Note 8). All samples were single phase as indicated by x-ray diffraction. After growth and cooling to room temperature, the $\sim$0.4 $\mu$m thick epilayers were {\it in situ} capped by an amorphous 50 nm thick Se layer, which was desorbed inside the ARPES and XMCD chambers by carefully annealing at $\sim 230\,^{\circ}{\rm C}$ for $\sim$1 h. The Mn concentrations determined by core-level photoemission were in good agreement with those obtained from energy dispersive microanalysis (EDX), indicating no significant Mn accumulation at the surface of the samples (see Supplementary Fig. 21 and Supplementary Note 9).\\ 

\noindent {\bf High-resolution and spin-resolved ARPES.}
Temperature-dependent ARPES measurements were performed at the UE112-PGM2a beamline of BESSY II at pressures better than $1\cdot10^{-10}$ mbar using p-polarized undulator radiation. Photoelectrons were detected with a Scienta R8000 electron energy analyzer and the spin-resolved spectra were obtained with a Mott-type spin polarimeter coupled to the hemispherical analyzer. For the spin-resolved measurements of Mn-doped \BiSe samples, a magnetic field of +0.3 T was applied perpendicular to the surface plane at 20 K right before the acquisition of the spectra. Overall resolutions of ARPES measurements were 5 meV (energy) and $0.3\,^{\circ}$ (angular). Resolutions for spin-resolved ARPES were 80 meV (energy) and $0.75\,^{\circ}$ (angular).\\

\noindent {\bf Magnetic characterization.}
The characterization of the bulk magnetic properties was done by SQUID magnetometry. The bulk magnetization was recorded as a function of temperature and applied magnetic field applied either perpendicular (out-of-plane) or parallel (in-plane) to the surface of the films. The diamagnetic contribution of the BaF$_2$ substrate was derived from the field-dependent magnetization curves at room temperature and subtracted from all data. The characterization of the surface magnetic properties was done by XMCD and XMCD-PEEM measurements at the UE46-PGM1 and UE49-PGM1a beamlines of BESSY II, respectively. The experiments were performed using circularly-polarized undulator radiation and under the same pressure conditions as the ARPES measurements. The XMCD absorption spectra were taken using a high-field diffractometer as a function of temperature and applied magnetic field, and the XMCD-PEEM measurements were performed at room temperature and under grazing incidence (16$^\circ$) with a lateral resolution of $\sim$20 nm.\\

\noindent {\bf Theoretical calculations.}
The one-step model photoemission calculations were performed using the results of {\it ab-initio} theory as an input, and taking into account wave vector and energy-dependent transition matrix elements. The {\it ab-initio} calculations were performed within the coherent-potential approximation using the KKR program package \cite{Ebert-KKR-2011}, with spin-orbit coupling included by solving the four-component Dirac equation.\\


\vspace{0.2in}
\noindent {\bf Acknowledgements}. Financial support from SPP 1666 of the Deutsche Forschungsgemeinschaft and the Impuls-und Vernetzungsfonds der Helmholtz-Gemeinschaft (Grant No. HRJRG-408) is gratefully acknowledged. J. M. is supported by the CENTEM (CZ.1.05/2.1.00/03.0088) and  CENTEM PLUS (LO1402) projects, co-funded by the ERDF as part of the Ministry of Education, Youth and Sports OP RDI programme. V. H. acknowledges the support of the Czech Science Foundation (project 14-08124S).\\

\noindent {\bf Author contributions}. J. S.-B and A. V performed photoemission experiments; G. S., G. B. and L. V. Y. provided samples and performed growth and characterization; H. S., R. K. and  A. N. performed SQUID measurements; O. C. and V. H. performed x-ray diffraction measurements; J. S.-B, E. S., E. W. performed XMCD measurements; J. S.-B, A. A. \"U. and S. V. performed XMCD-PEEM measurements; J. M., M. D., J. B and H. E. carried out calculations; J. S.-B and E. G. performed numerical simulations; J. S.-B performed data analysis and figure planning; J. S.-B. and O. R. performed draft planning and wrote the manuscript with input from all authors; J. S.-B. and O. R. were responsible for the conception and the overall direction.\\

\noindent {\bf Competing financial interests}. The authors declare no competing financial interests. Correspondence and requests for materials should be addressed to J. S.-B. (Email: jaime.sanchez-barriga@helmholtz-berlin.de) and O. R. (Email: rader@helmholtz-berlin.de).\\

\newpage

\noindent {\bf FIGURES AND CAPTIONS}\\

\phantom{Zeile}

\begin{figure}[h]
\centering
\includegraphics [width=0.9\textwidth]{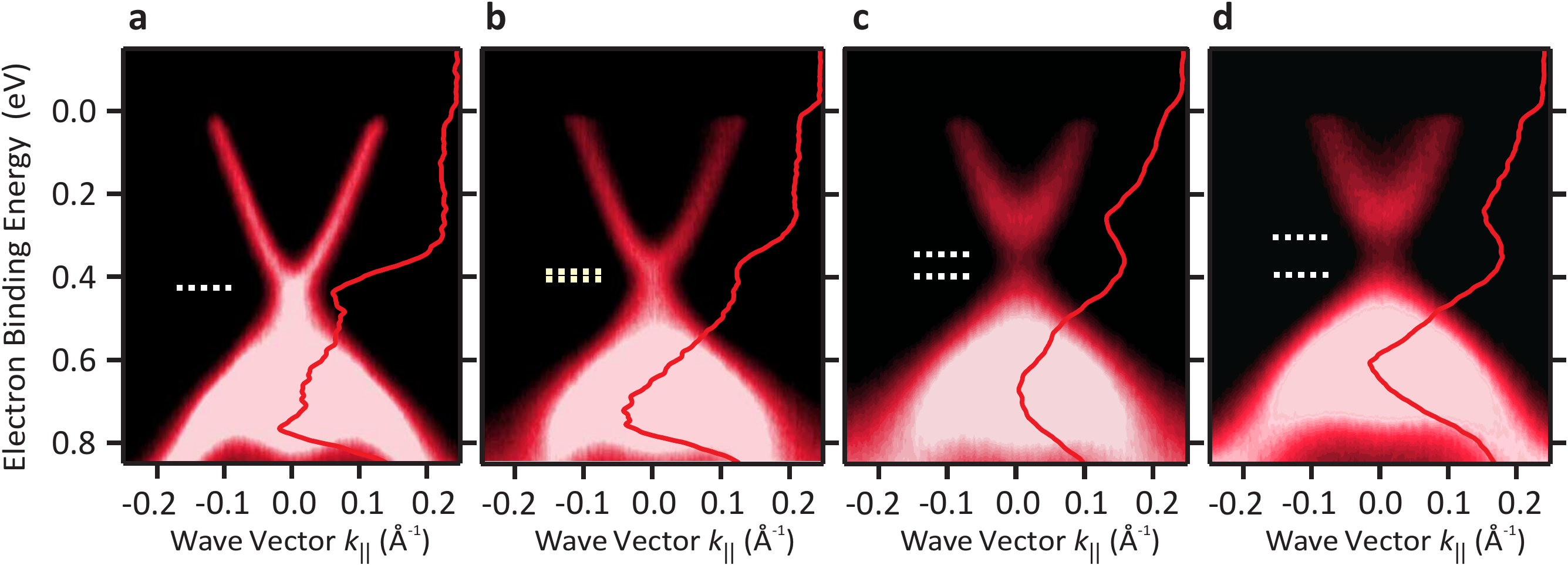}
 \caption{ {\bf Effect of Mn doping on the electronic band dispersions of \BiMnSe}. (a-d) Mn-concentration dependent high-resolution ARPES dispersions measured at 50 eV photon energy and a temperature of 12 K for $x$ values of (a) 0, (b) 0.02, (c) 0.04 and (d) 0.08. The red lines represent energy-distribution curves (EDCs) obtained in normal emission (${k}_\parallel$=0). A surface band gap opens at the Dirac point of the TSS which increases in size with increasing Mn content. The horizontal white-dashed lines highlight the minimum limit of the gap size after taking into account the contribution from the linewidth broadening. The opening of the gap leads to an intensity dip in the EDCs around the energy region of the original Dirac point. For $x=0.02$, the suppression of intensity at the Dirac point and the more parabolic surface-state dispersion are signatures of a small gap. Besides the increased linewidth broadening due to the Mn impurities, these effects become more pronounced with increasing Mn content.}
\label{fig1}
\end{figure}

\phantom{Zeile}

\begin{figure}[h]
\centering
\includegraphics [width=0.65\textwidth]{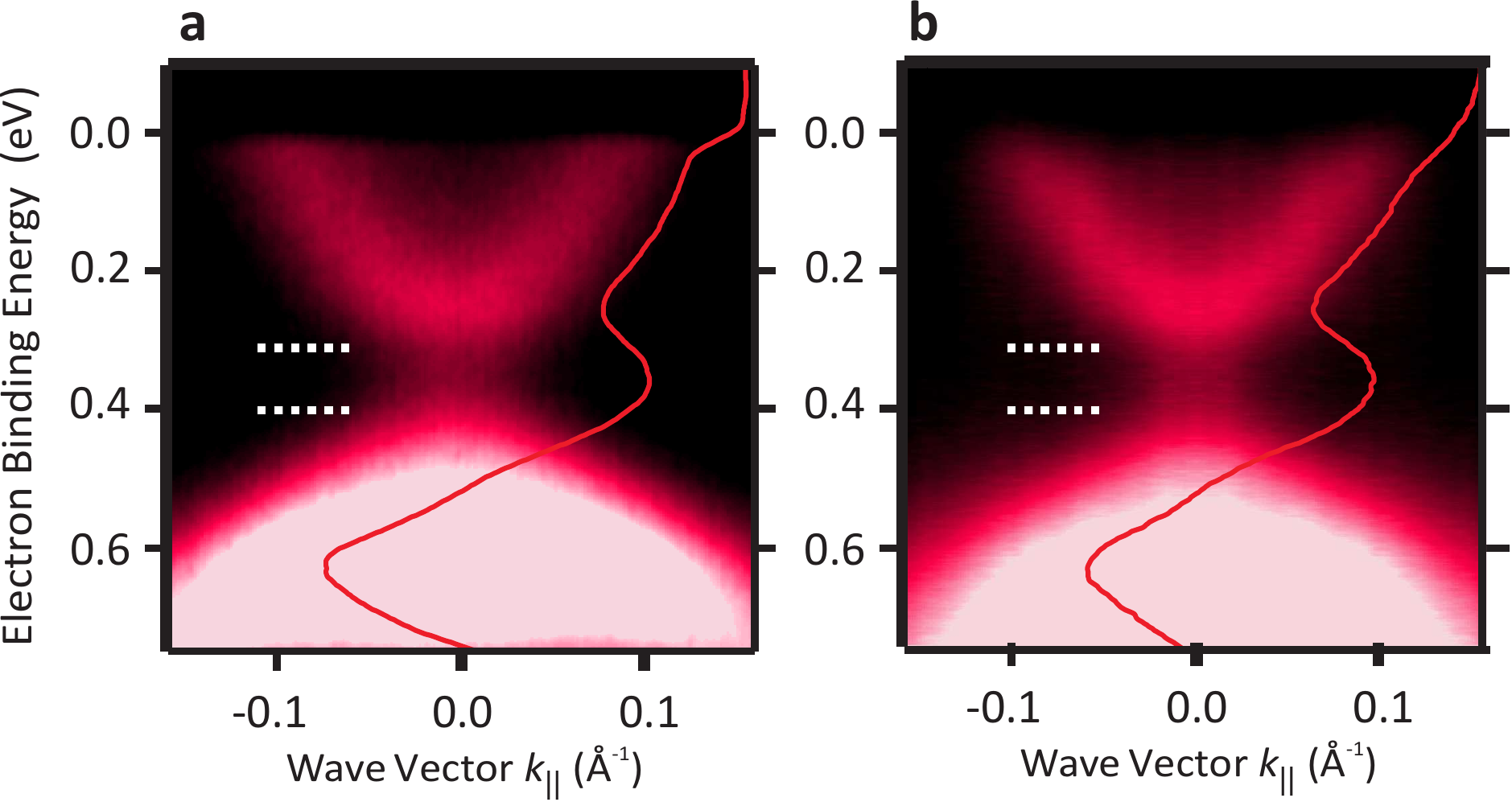}
\caption{ {\bf Temperature dependence of the large energy scale surface band gap}. (a, b) Energy-momentum ARPES dispersions obtained for Mn-doped \BiSe\ epilayers with 8\%\ Mn concentration. The red lines are the energy-distribution curves obtained in normal emission (${k}_\parallel$=0). The horizontal white-dashed lines highlight the surface gap. Measurements are taken at 50 eV photon energy and a temperature of (a) 12 K and (b) 300 K. The surface band gap does not show a remarkable temperature dependence.}
\label{fig2}
\end{figure}

\phantom{Zeile}

\begin{figure}
\centering
\includegraphics [width=0.8\textwidth]{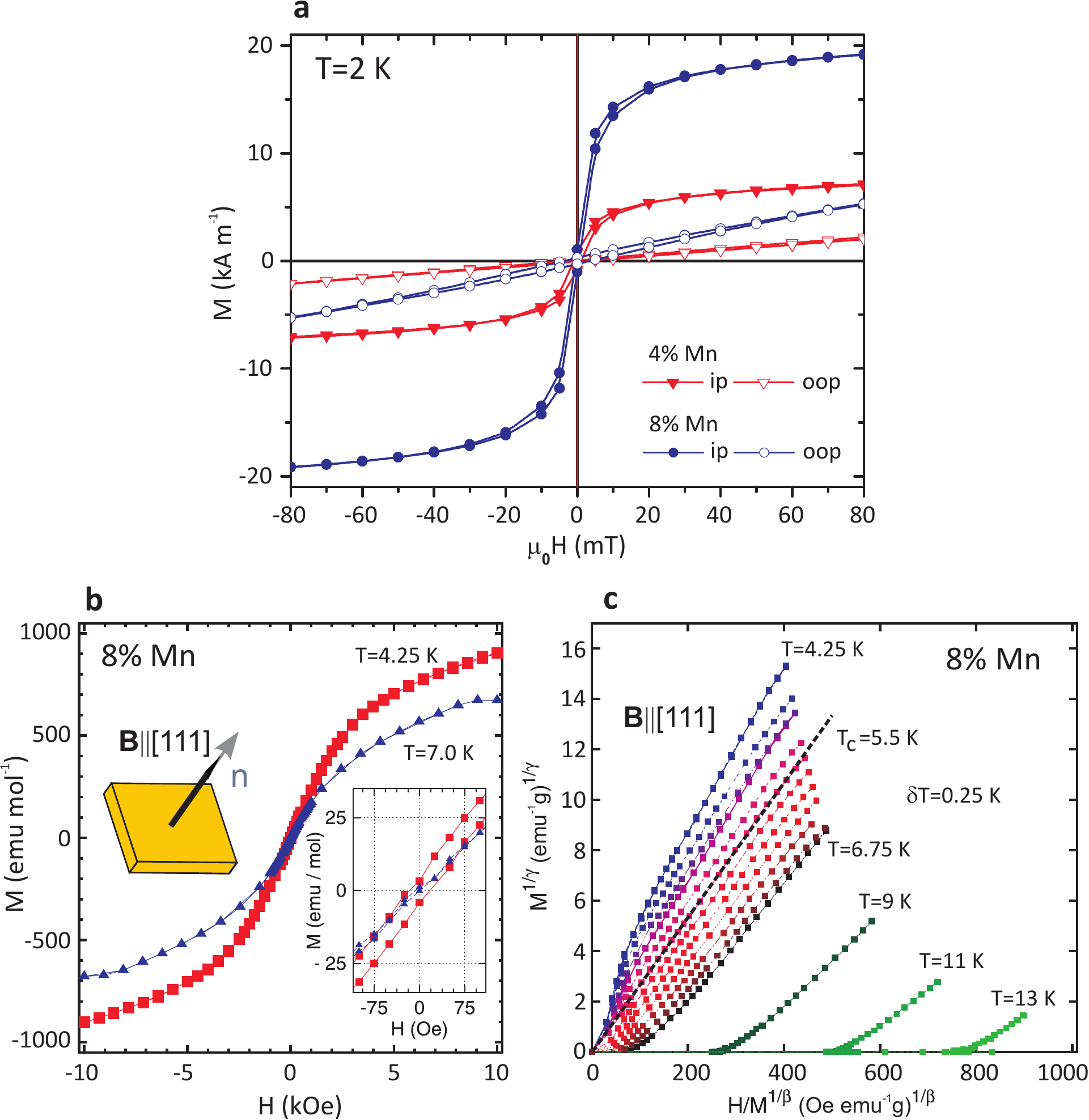}
\caption{{\bf Characterization of the bulk magnetic properties of Mn-doped \BiSe\ films}. (a) In-plane (ip) and out-of-plane (oop) magnetization curves for Mn concentrations of $x=0.04$ and $x=0.08$, measured at a temperature of 2 K using SQUID magnetometry. The magnetic anisotropy lies in the surface plane irrespective of the Mn concentration. (b) Corresponding hysteresis loops for $x=0.08$ measured at a temperature of 4.25 K (squares) and 7 K (triangles), respectively. The applied magnetic field is perpendicular to the (111) sample surface (see sketch). Inset: A zoom-in around zero magnetic field showing hysteresis at 4.25 K and a paramagnetic state at 7 K. (c) Arrott plots at various temperatures from which a Curie temperature $T_{\rm C}$ of 5.5 K is deduced. Between 4.25 and 6.75 K, data are shown for temperature increments of $\delta T=0.25$ K.}
\label{fig3}
\end{figure}

\phantom{Zeile}

\begin{figure}
\centering
\includegraphics [width=0.7\textwidth]{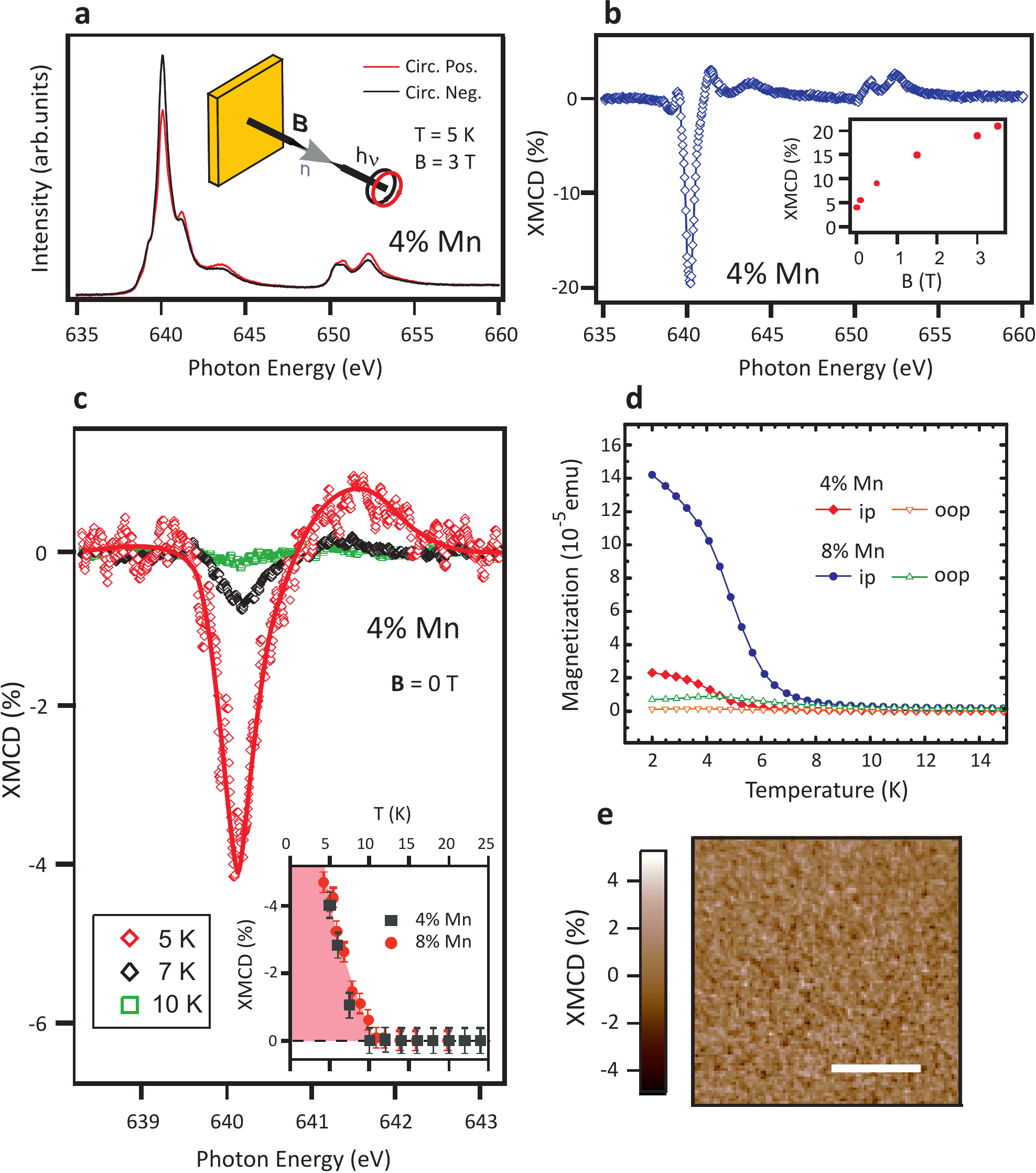}
\caption{{\bf XMCD in total electron yield at the Mn $L_{2,3}$-edges}. (a) X-ray aborption spectra measured for opposite helicities of the circularly-polarized incident light at 5 K with an out-of-plane applied magnetic field of 3 T. A sketch of the experimental geometry is also shown. (b) Corresponding XMCD difference spectrum. The inset shows XMCD measurements for different applied magnetic fields. The error bars are estimated from the average noise level of the XMCD curves. (c) Temperature dependence of the remanent XMCD for a Mn concentration of $x=0.04$. Inset: Detailed temperature dependence for $x=0.04$ and $0.08$. (d) To re-examine the bulk magnetic properties, field-cooling measurements are obtained under an in-plane (ip) and out-of-plane (oop) applied magnetic field of 10 mT using SQUID magnetometry. The temperature dependence for $x=0.04$ and $0.08$ compares well to the surface-sensitive measurements shown in the inset of (c). (e) XMCD-PEEM image obtained using photoelectron microscopy for $x=0.08$ at room temperature, revealing the absence of magnetic domains with partial or full out-of-plane magnetic order. The scale bar (horizontal white-solid line) is 500 $\mu$m.}
\label{fig4}
\end{figure}

\phantom{Zeile}

\begin{figure}[t]
\centering
\includegraphics [width=0.8\textwidth]{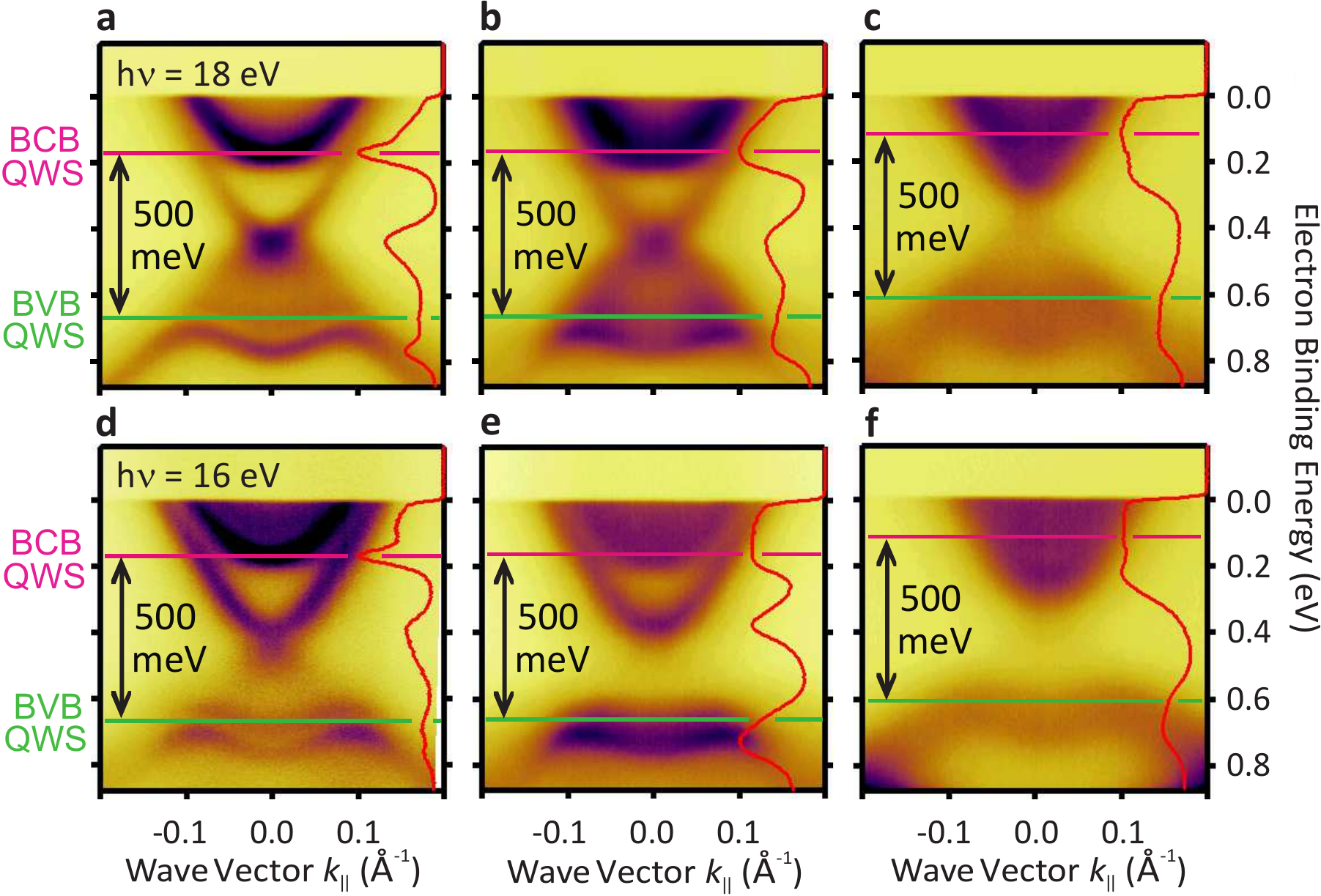}
 \caption{{\bf Tracing changes in the bulk band gap of Mn-doped \BiSe\ films}. Changes in the bulk band gap are traced most precisely from quantum-well states (QWS) in normal emission (red curves on the right-hand side of each panel).(a-c) Energy-momentum dispersions acquired at a photon energy of $h\nu=18$ eV for (a) \BiSe, (b) 2\% and (c) 8\% Mn doping. The simultaneous quantization of bulk-conduction band (BCB) and valence band (BVB) is created by surface band bending after adsorption of small amounts of residual gas \cite{BianchiPRL11}. It is found that Mn doping does not change the bulk band gap of \BiSe. (d-f) Similar results as in (a-c), respectively, but at a photon energy of $h\nu = 16$ eV, where large apparent surface band gaps are observed. This unexpected photon-energy dependence of a surface state is interpreted as coupling to bulk states mediated by the Mn impurities.}
\label{fig5}
\end{figure}

\phantom{Zeile}

\begin{figure}
\centering
\includegraphics [width=0.65\textwidth]{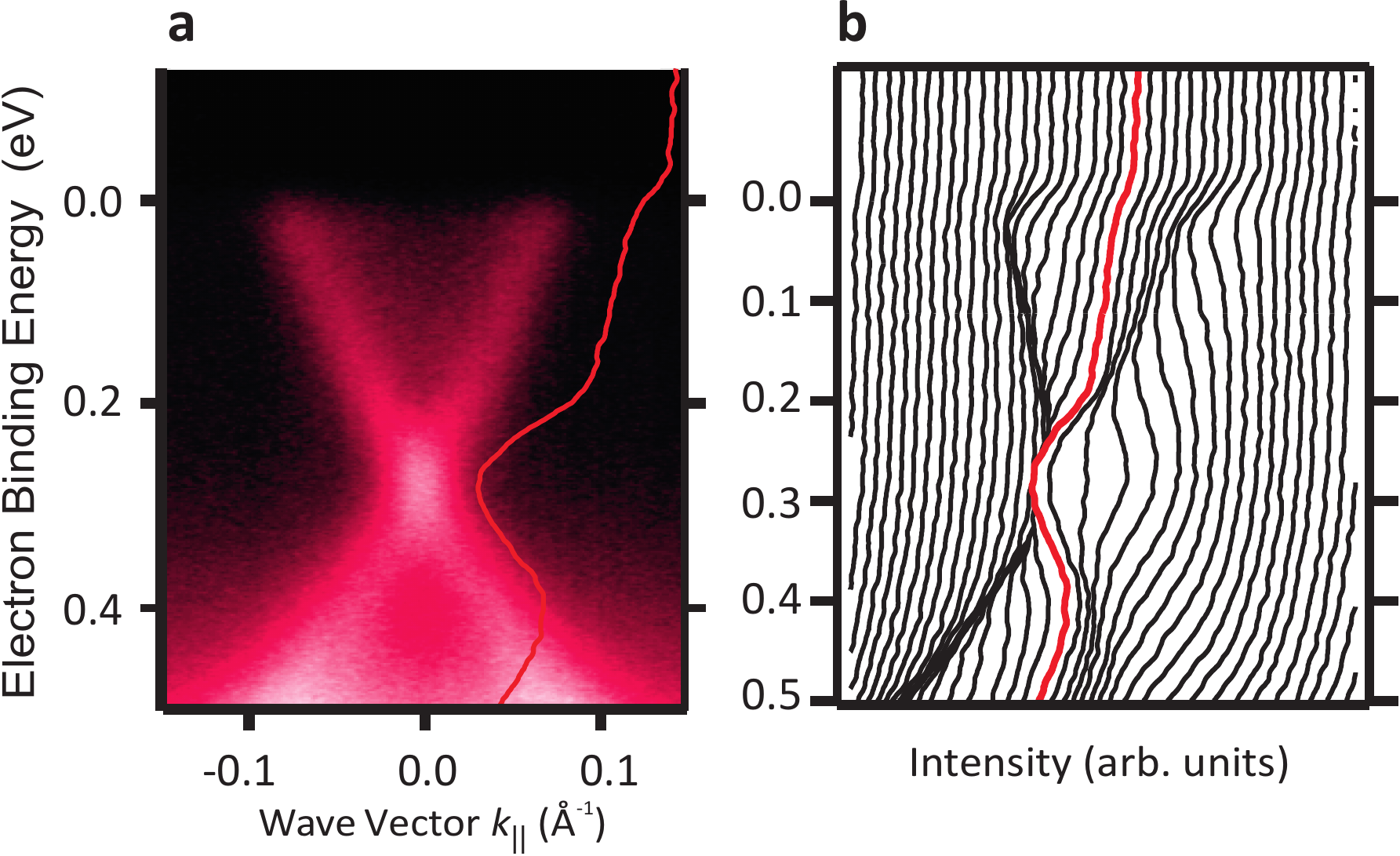}
 \caption{{\bf Effect of Mn deposition on the surface of \BiSe\ films}. (a) Energy-momentum ARPES dispersion showing a gapless Dirac cone in pure \BiSe\ films after deposition of 0.3 monolayer Mn at 30 K on the surface. Measurements are performed at the same temperature using 50 eV photon energy. (b) Energy-distribution curves (EDCs) extracted from the measurements shown in (a). The red curves in (a) and (b) emphasize the EDC in normal emission (${k}_\parallel$=0).}
\label{fig6}
\end{figure}

\phantom{Zeile}

\begin{figure}
\centering
\includegraphics [width=0.5\textwidth]{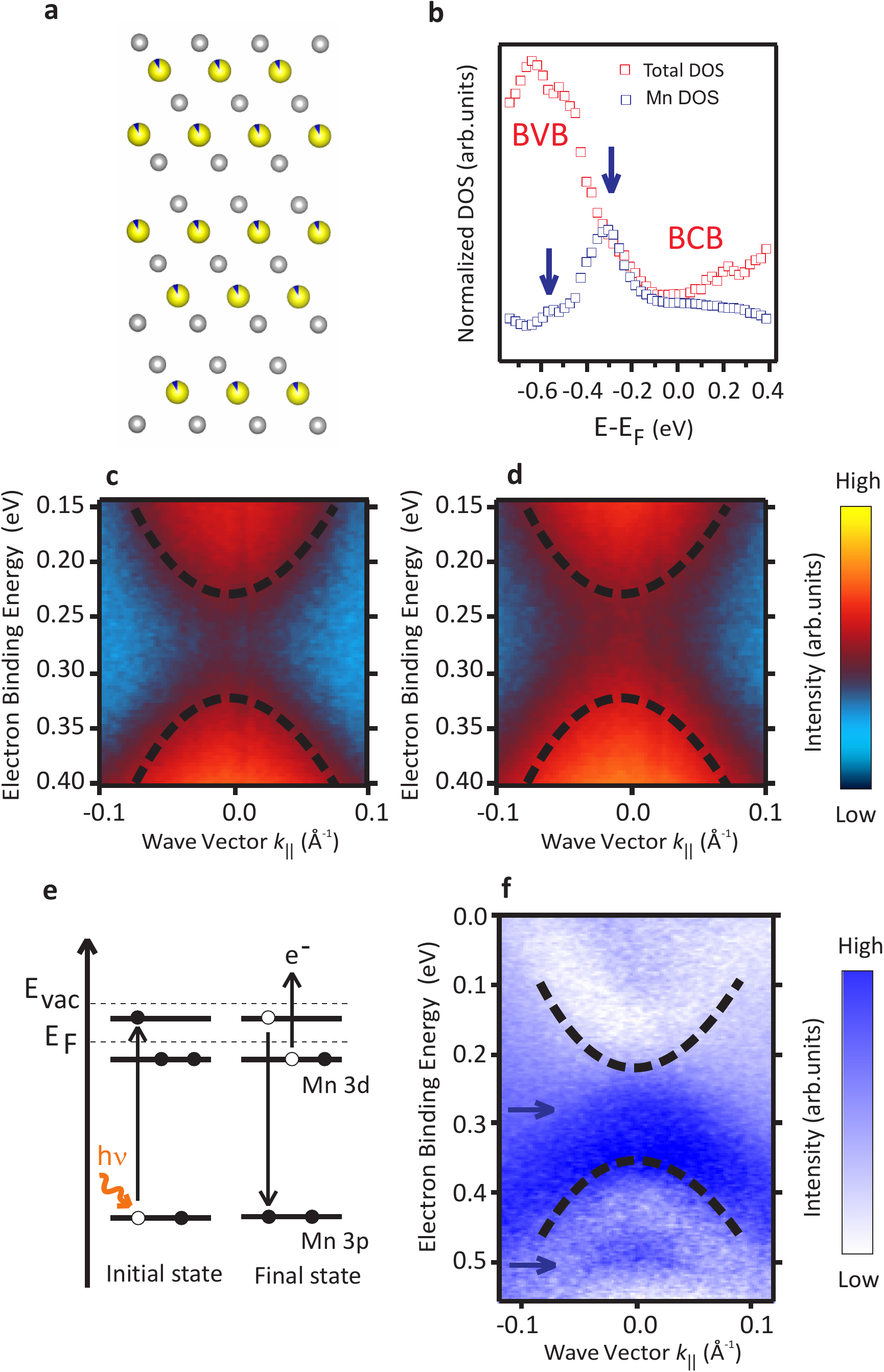}
\caption{{\bf Resonant photoemission and ab-initio calculations}. (a) The structure of \BiSe\ doped with 8\% Mn at the Bi-substitutional sites used in the calculations. Within the coherent-potential approximation, the Bi atoms (yellow) acquire Mn character (blue). Se atoms are shown in gray color. (b) Calculated density of states (DOS). The total DOS (red) contains contributions from impurity resonances (vertical blue arrows) of strong d-character, a seen in the partial Mn DOS (blue). For the purpose of the calculation, ferromagnetic order at $T=0$ K was assumed. (c, d) Resonant photoemission at the Mn $M$-edge, focusing on the region of the surface gap (8\% Mn). (c) Off-resonant (48 eV photon energy) and (d) on-resonant data (50 eV). (e) Schematics of the resonant process. At the resonance energy, the excitation occurs via 3$p$-3$d$ transitions. The exciting photon energy (h$\nu$) corresponds to a transition from an occupied core level to a valence band empty state between the Fermi (E$_F$) and vacuum (E$_{vac}$) levels. The relaxation of excited electrons leads to enhanced photoemission from $d$-like Mn states. (f) Intensity difference obtained after subtracting off-resonant from on-resonant data. The resonances are seen in blue contrast and are marked with arrows, similar to the calculations shown in (b).}
\label{Fig7}
\end{figure}

\end{document}